\documentclass[10pt]{article}

\usepackage{amsmath,amssymb,amsfonts,amsthm}
\usepackage{color,graphicx}
\usepackage{psfrag}

\newcommand{\vw}{{\mathbf{w}}}

\newcommand{\cT}{{\mathcal T}}
\newcommand{\cQ}{{\mathcal Q}}

\newcommand{\dt}{\partial_t}
\newcommand{\dx}{\partial_x}
\newcommand{\eps}{\varepsilon}
\newcommand{\dsp}{\displaystyle}

\newcommand{\hm}{\hspace{-0.5em}}

\newcommand{\beq}{\begin{equation}}
\newcommand{\eeq}{\end{equation}}
\newcommand{\be}{\begin{equation}}
\newcommand{\ee}{\end{equation}}
\newcommand{\bes}{\begin{equation*}}
\newcommand{\ees}{\end{equation*}}
\newcommand{\bea}{\begin{eqnarray}}
\newcommand{\eea}{\end{eqnarray}}
\newcommand{\beas}{\begin{eqnarray*}}
\newcommand{\eeas}{\end{eqnarray*}}
\newcommand{\beqn}{\begin{eqnarray}}
\newcommand{\eeqn}{\end{eqnarray}}
\newcommand{\beqns}{\begin{eqnarray*}}
\newcommand{\eeqns}{\end{eqnarray*}}

\newtheorem{thrm}{Theorem}[section]

\newtheorem{remark}[thrm]{Remark}

\numberwithin{equation}{section}
\begin{document}

\title{Numerical simulation of strongly nonlinear and dispersive waves  using a Green-Naghdi model}
\date{}
\author{F. Chazel$^{1}$, D. Lannes$^2$ and F. Marche$^3$\\[2pt]
$^1$ Universit\'e de Toulouse, UPS/INSA, IMT, CNRS UMR 5219,\\
Florent.Chazel@math.univ-toulouse.fr\\[2pt]
$^2$DMA, Ecole Normale Sup\'erieure, et CNRS UMR 8553\\
David.Lannes@ens.fr\\[2pt]
$^3$I3M, Univ. Montpellier 2 et CNRS UMR 5149\\
Fabien.Marche@math.univ-montp2.fr}

\maketitle

\begin{abstract}
We investigate here the ability of a Green-Naghdi model to reproduce strongly nonlinear and dispersive wave propagation. We test in particular the behavior of the new hybrid finite-volume and finite-difference splitting approach recently developed by the authors and collaborators   on the challenging benchmark of waves propagating over a submerged bar. Such a configuration requires a model with very good dispersive properties, because of the high-order harmonics generated by topography-induced nonlinear interactions.  We thus depart from the aforementioned work and choose to use  a new Green-Naghdi system with improved frequency dispersion characteristics. The absence of dry areas also allows us to improve the treatment of the hyperbolic part of the equations. This leads to very satisfying results for the demanding benchmarks under consideration.
\end{abstract}

\section{Introduction}

The propagation of surface waves in an incompressible, homogeneous, inviscid fluid is governed by the free surface Euler equations. In shallow water (i.e. when the typical wavelength of the waves is much larger than the typical depth), a good alternative is furnished by the Green-Naghdi equations (also called Serre or fully nonlinear Boussinesq equations); these equations were first derived in \cite{sug1969, gre1976, sea1987} (see also \cite{cie2006,wei1995}), and we refer to \cite{Lannes-Bonneton} for a recent review and a derivation  under the formulation used here. It is known \cite{AL} that they approximate the full Euler equations with a good accuracy up to the breaking point.

\medbreak

Let $h_0$ denote the reference depth, $\zeta(t,x)$ the elevation of the free surface with respect to its rest state, and $b(x)$ the variation of the bottom topography with respect to the constant level $z=-h_0$. The Green-Naghdi equations couple the evolution of the water depth $h(t,x)=\zeta(t,x)+h_0-b(x)$ to the evolution of $u(t,x)$, the vertically averaged horizontal component of the velocity. For $1d$ surface waves, these equations can be written
\be\label{GNdim}
\!\!
	\left\lbrace
	\begin{array}{l}
	\vspace{0.5em}
	\!\!\! \dt h + \dx (h u)=0,\\
       	 \!\!\!\dt (hu) + \dx(hu^2) + \frac{\alpha-1}{\alpha} gh\dx\zeta + (1\!+\!\alpha h\cT\frac{1}{h})^{-1}
        [\frac{1}{\alpha}gh\dx\zeta + h\cQ(u)] \!=\! 0,
	\end{array}\right.
\ee
where $\alpha\geq 1$ and the operators $\cT$ and $\cQ$ are explicitly given by
\be \label{defT}
\cT w = -\frac{h^2}{3} \partial_x^2 w - h \partial_x h \partial_x w + (\partial_x \zeta \partial_x b + \frac{h}{2} \partial_x^2 b) w,
\ee
and
\be \label{defQ}
\cQ(u) = 2h\partial_x (h+\frac{b}{2}) (\partial_x u)^2 + \frac{4}{3} h^2 \partial_x u \partial_x^2 u + h \partial_x^2 b u \partial_x u
+ (\partial_x \zeta \partial_x^2 b + \frac{h}{2} \partial_x^3 b) u^2.
\ee
\begin{remark}\label{remalpha}
The usual Green-Naghdi equation corresponds to $\alpha=1$ in (\ref{GNdim}). Using the classical
observation that
\begin{eqnarray*}
\dt u&=&-\dx\zeta-u\dx u+\mbox{higher order terms}\\
&=&\alpha\dt u-(1-\alpha)(\dx\zeta+u\dx u)+\mbox{higher order terms},
\end{eqnarray*}
one deduces (\ref{GNdim}), with the same accuracy (see \cite{BCLMT} for details). 
\end{remark}

Looking at the linearization of (\ref{GNdim}) around the steady state $h=h_0$, $u=0$ over a flat bottom $b=0$, one derives
the dispersion relation associated to (\ref{GNdim}). It is found  by looking for plane wave solutions of the 
form $(\underline{h},\underline{h}\underline{u})e^{i(k x-\omega t)}$ to the linearized equations, and is given by,
\begin{equation}\label{reldispGN}
\omega_{\alpha}^2(k)= gh_0 k^2 \frac{1+(\alpha-1)(k h_0)^2/3}{1+\alpha(k h_0)^2/3}.
\end{equation}
The interest of the parameter $\alpha$ introduced in Remark \ref{remalpha} is that it can be chosen in such a way that (\ref{reldispGN}) adequately matches the exact dispersion relation $\omega^2(k)=gk\tanh(kh_0)$ of the full Euler equation, even for non-small values of $kh_0$. Taking $\alpha=1.159$ as in \cite{BCLMT} yields a good agreement up to $kh_0=4$ on the linear phase velocity and up to $kh_0=2.5$ on the linear group velocity. 

\medbreak

The goal of \cite{BCLMT} is to develop a numerical code having good dispersive properties, and able to handle successfully wave breaking and vanishing depth (shoreline). The idea is to use a splitting scheme that decomposes the dispersive and nonlinear parts of the equations.  More precisely, using a second order splitting scheme,  the approximation $U^{n+1}=(\zeta^{n+1},u^{n+1})$ at time $(n+1)\delta_t$ 
is found in
terms of the approximation $U^n$ at time $n\delta_t$ by solving
$$
U^{n+1}=S_1(\delta_t/2)S_2(\delta_t)S_1(\delta_t/2)U^n,
$$
where $S_1(\cdot)$ is the solution operator associated to the nonlinear shallow water (or Saint-Venant) equations  and
$S_2(\cdot)$ the solution operator associated to the dispersive effects. It is shown in \cite{BCLMT} how to adapt this method to deal with wave breaking and vanishing depth. Validations relying on analytical solutions and comparisons with experimental data give very good results. For most of the applications, the dispersive properties of this model (with $\alpha=1.159$ in (\ref{GNdim})) are good enough. In order to consider challenging configurations that include high harmonics (typically up to $kh_0=4$), we first derive a new family of Green-Naghdi models with improved frequency dispersion following the steps of Nwogu and adapt the approach of \cite{BCLMT} to this new model. Since this is not relevant for this test case, we do not mention here the treatment of wave breaking nor of the shoreline (see \cite{BCLMT} for details); on the other hand we propose numerical improvements on the computation of the hyperbolic part made possible by the absence of dry areas.

\medbreak

The paper is organized as follows. In Section \ref{secttheta}, we derive a new family of Green-Naghdi equations that have the same precision as the original one as $kh_0 \to 0$, but with better dispersive properties. We then present in Section \ref{sectNS} the main features of the numerical scheme. Finally, we present in Section \ref{NV} some numerical validations of our derived models.

\section{A three-parameters family of GN models}\label{secttheta}

The main goal of this section is to derive a new family of Green-Naghdi equations that are equivalent to the original set, up to higher order terms. As a mean to clearly identify these higher order terms in our fully nonlinear and weakly dispersive framework, we rather work on the nondimensionalized version of (\ref{GNdim}), namely,  
\be\label{GNadim}
	\left\lbrace
	\begin{array}{l}
	\vspace{0.5em}
	\!\!\!\dt h + \dx (h u)=0,\\
	\vspace{0.5em}
       	\!\!\!(1 \!+\! \mu\alpha h\cT\frac{1}{h}) \big[\dt(hu) + \eps \dx(hu^2) + \frac{\alpha-1}{\alpha} h\dx\zeta\big]
         \!+ \!\frac{1}{\alpha} h\dx\zeta + \varepsilon\mu h Q(u) = 0,
	\end{array}\right.
\ee
where $\eps = a/h_0 = O(1)$ is the nonlinearity parameter and $\mu = h_0^2/\lambda^2 \ll 1$ is the shallowness parameter, $a$ being the typical wave and bottom deformation amplitude, $\lambda$ the typical wavelength and $h_0$ the reference depth. We still denote by $h$ the dimensionless water depth, $h=1+\eps(\zeta-b)$.

\medbreak

In \cite{Nwogu}, Nwogu showed that it was possible to improve the dispersive properties of Boussinesq models by working with the velocity at a certain depth as dependent variable. This approach was generalized in \cite{wei1995} to the fully nonlinear case (Green-Naghdi). When the bottom is not flat, it turns out that in the fully nonlinear case, the standard Green-Naghdi equations written with the average velocity $u$ do \emph{not} belong to this new class of fully nonlinear models. This is the reason why we use a slightly different approach here, with the introduction of a new dependent variable $u_\theta$ that is not the velocity at a certain depth. The interest is that the computations are somehow simpler and, more importantly, that the average velocity $u$ appears as a particular case ($\theta=0$) --- our choice of dependent variables also differs from the one used in \cite{GN_Grav} and whose purpose is not to improve the dispersive properties but to work with potential variables.\\
For all $\theta \geq 0$, we thus define 
\be\label{uth}
u_{\theta}=(1+\mu\theta \cT)^{-1}u.
\ee
The first equation of (\ref{GNadim}) can thus be written
\begin{equation}\label{first}
\dt h+\dx (h u_\theta)+\mu\theta\dx (h\cT u_\theta)=0.
\end{equation}
Note that this equation, as well as the first equation of (\ref{GNadim}) is \emph{exact} (and not up to higher order terms). For the second equation, the transformation is more complex. Let us first focus on the component $(1+\mu\alpha h \cT \frac{1}{h})\dt (hu)$. Using (\ref{uth}) we can write
\begin{equation}
\dt (hu) = (1+\mu\theta h\cT\frac{1}{h})\dt (hu_\theta)+\mu\theta [\dt,h\cT\frac{1}{h}]h u_\theta\label{uth1}.
\end{equation}
Using (\ref{defT}), we deduce after some computations that 
\begin{eqnarray*}
[\dt,h\cT\frac{1}{h} ]w \hm& = &\hm -\frac{2}{3}h\dt h\dx^2 w-\frac{1}{3}\dx (h\dt h)\dx w\\
\hm& &\hm +\big( \frac{1}{3}\dx^2(h\dt h)+\dx (\dt h)\dx b+\frac{1}{2}(\dt h)\dx^2 b\big)w;
\end{eqnarray*}
with the help of (\ref{first}), this shows that  (\ref{uth1}) can be put under the form
\begin{equation}\label{pouet1}
\dt (hu)=\big(1+\mu\theta h \cT \frac{1}{h}\big)\dt (hu_\theta)+\eps\mu\theta \cQ_1(hu_\theta)+O(\mu^2),
\end{equation}
with
\be\label{Q1}
\cQ_1(m)=\frac{1}{2}h\dx (m_x^2)+\frac{1}{3}(\dx h) m_x^2
-\frac{1}{3}m\dx^2(h m_x) - \big(m_{xx}\dx b+\frac{1}{2}m_x\dx^2 b\big)m.
\ee
Let us now focus on the component $\eps(1+\mu\alpha h\cT\frac{1}{h})\dx (hu^2)$. Using (\ref{uth}), one has
\begin{eqnarray*}
\eps\dx (hu^2) & = & \eps \dx (hu_\theta^2) + \eps\mu\theta \dx (2 h u_\theta \cT u_\theta)+O(\mu^2),\nonumber\\
               & = & \eps (1+\mu\theta h\cT\frac{1}{h}) \dx (hu_{\theta}^2) + \eps\mu\theta \cQ_r(u_\theta) + O(\mu^2),
\end{eqnarray*}
with $\cQ_r = \dx (h(2u_\theta\cT u_\theta - \cT u\theta^2)) + [\dx, h\cT\frac{1}{h}](hu_\theta^2)$.
After some computations, we obtain
$
\cQ_r(u_\theta) = h \cQ(u_\theta) + \cQ_2(u_\theta)
$
with
\begin{eqnarray}\label{Q2}
\cQ_2(u_\theta) \hm& = &\hm -\frac{1}{3}\dx(h^2)\partial_x^2(hu_\theta^2) - \Big(\frac{1}{6}\partial_x^2(h^2)-\eps\dx\zeta\dx b\Big)\dx(hu_\theta^2)
- h^2\dx b (\dx u_\theta)^2 \nonumber \\
\hm& &\hm + \Big(\frac{1}{6}\partial_x^3(h^2) + \dx (2\eps\dx\zeta\dx b+\frac{h}{2}\partial_x^2 b) - \dx b \partial_x^2 b \Big) h u_\theta^2.
\end{eqnarray}
We have thus proved that
\be\label{uth2}
\eps\dx (hu^2)=\eps (1+\mu\theta h\cT\frac{1}{h})\dx (hu_\theta^2)+\eps\mu\theta\big(h\cQ(u_\theta)+\cQ_2(u_\theta)\big)+O(\mu^2).
\ee
Pluging  (\ref{pouet1}),(\ref{uth2}) into the second equation of (\ref{GNadim}) and recalling (\ref{first}), we thus get that (\ref{GNadim}) is equivalent, up to $O(\mu^2)$ terms in the second equation, to
\begin{equation}\label{GNtheta_adim}
\left\lbrace
\begin{array}{l}
 \dt h + \eps\dx (hu_\theta)+\eps\mu\theta\dx (h\cT u_\theta)=0,\vspace{1mm}\\
(1+\mu(\alpha+\theta)h\cT\frac{1}{h})\big[\dt (hu_\theta) + \eps\dx (hu_\theta^2) + \frac{\alpha-1}{\alpha+\theta} h\dx\zeta\big]  \\
\hspace{0.5cm} + \frac{1+\theta}{\alpha+\theta} h \dx\zeta+\eps\mu\big((1+\theta)h Q(u_\theta) + \theta (Q_1(h u_\theta) + Q_2(u_\theta))\big)=0.
\end{array}\right.
\end{equation}
At this step, we point out that the introduction of the parameter $\theta$ reveals to be useless as far as the minimization of the dispersion errors is concerned. Indeed, keeping in mind that $\alpha \ge 1$ and $\theta \ge 0$, computing the dispersion relation associated to these new equations and minimizing the dispersion errors (on the phase and group velocities) over the range $kh \in [0,4]$ yields the values $\alpha = 1.159$ and $\theta = 0$.\\
A workaround for this issue is to introduce a third parameter $\gamma \ge 0$ by applying the operator $(1+\mu\gamma h\cT\frac{1}{h})$ to the first equation of (\ref{GNtheta_adim}). Doing so, the first non-dimensional equation becomes
$$
\dt h + \eps\dx (hu_\theta) + \eps\mu\theta (1+\mu\gamma h\cT\frac{1}{h})^{-1} \dx (h\cT u_\theta)=O(\mu^2).
$$
The drawback of this manipulation is that the exactness of the equation is lost; nevertheless, the numerical computations show that it is a fair price to pay for the considerable improvement of the dispersive properties thus achieved.\\
On the whole, this final manipulation yields a new three-parameter family $(\mathcal{G})$ of Green-Naghdi models $\mathcal{G}_{\alpha,\theta,\gamma}$ that can be written in dimensional form as
\begin{equation}\label{GNgamma}
\mathcal{G}_{\alpha,\theta,\gamma}
\left\lbrace
\begin{array}{l}
\displaystyle \dt h+\dx (hu_\theta)+\theta (1+\gamma h\cT\frac{1}{h})^{-1} \dx (h\cT u_\theta)=0,\vspace{1mm}\\
\displaystyle\dt (hu_\theta) + \dx (hu_\theta^2) + \frac{\alpha-1}{\alpha+\theta} gh\dx\zeta\\
\hspace{1cm}\displaystyle+(1+(\alpha+\theta)h\cT\frac{1}{h})^{-1}\Big(\frac{1+\theta}{\alpha+\theta} gh \dx\zeta+h\widetilde{\cQ}(u_\theta)\Big)=0,
\end{array}\right.
\end{equation}
with $\alpha \ge 1$, $\theta \ge 0$ and $\gamma \ge 0$ and
\begin{equation}\label{Qbis}
\widetilde{\cQ}(u_\theta)=(1+\theta)\cQ(u_\theta)+\frac{\theta}{h}\big(\cQ_1(hu_\theta)+\cQ_2(u_\theta)\big).
\end{equation}
\begin{remark}
We insist on the fact that all these models are equivalent at order $O(\mu^2)$ to the original Green-Naghdi equations (\ref{GNdim}), which  correspond to the particular case $\alpha=1$, $\theta=\gamma=0$. 
%Under this formulation, they share with (\ref{GNdim}) the interesting feature that no third order derivative of $\zeta$ is needed, which is interesting from the numerical viewpoint since this term can become quite stiff when the wave gets steeper.
\end{remark}

Now, one easily computes the dispersion relation associated to (\ref{GNgamma}),
$$
\dsp \omega_{\alpha,\theta,\gamma}^2(k)=gh_0k^2\frac{\big(1+\frac{\theta+\gamma}{3}(kh_0)^2\big)\big(1+\frac{\alpha-1}{3}(kh_0)^2\big)}
{\big(1+\frac{\gamma}{3}(kh_0)^2\big)\big(1+\frac{\alpha+\theta}{3}(kh_0)^2\big)}.
$$
Starting from this expression, one can deduce the expression - not reported here - of the associated linear phase and group velocities, denoted $C_{GN}^{p}$ and $C_{GN}^{g}$, along with the linear shoaling coefficient $\gamma_s$. Using these expressions, one computes for each triplet $(\alpha,\theta,\gamma)$ the corresponding dispersive errors between these model properties and the theoretical ones coming from Stokes theory, namely $C_{S}^{p}$, $C_{S}^{g}$ and $\gamma_0$ : see \cite{cie2007,FC_DL} for further details. The minimization of the shoaling error is quite problematic as a better shoaling involves a deterioration of the phase and group velocities, and vice-versa. For this reason, we provide here two different sets of optimized triplet $(\alpha,\theta,\gamma)$, one corresponding to optimal phase and group velocities errors only - which can be seen as optimal values for flat bottoms - and one corresponding to optimal phase and group velocities and shoaling errors. Optimizing over the dispersive range $kh_0 \in [0,4]$ yields :
\be\label{opt}
\alpha_{opt} = 1.028,\quad\theta_{opt} = 0.188,\quad\gamma_{opt} = 0.112,\quad\mbox{for flat bottoms},
\ee
and
\be\label{opt2}
\alpha_{opt}^{\,b} = 1,\quad\theta_{opt}^{\,b} = 0.207,\quad\gamma_{opt}^{\,b} = 0.071,\quad\mbox{for uneven bottoms}.
\ee

\begin{figure}[ht!]
\label{disp}
\psfrag{y}[cr][cl][0.8]{$\dsp\frac{C_{GN}^{p}}{C_{S}^{p}}$}
\psfrag{Linear phase velocity error}[0.7]{\hspace{7em}{\small Linear phase velocity error}}
\hspace{0.5em}\includegraphics[width=0.985\textwidth]{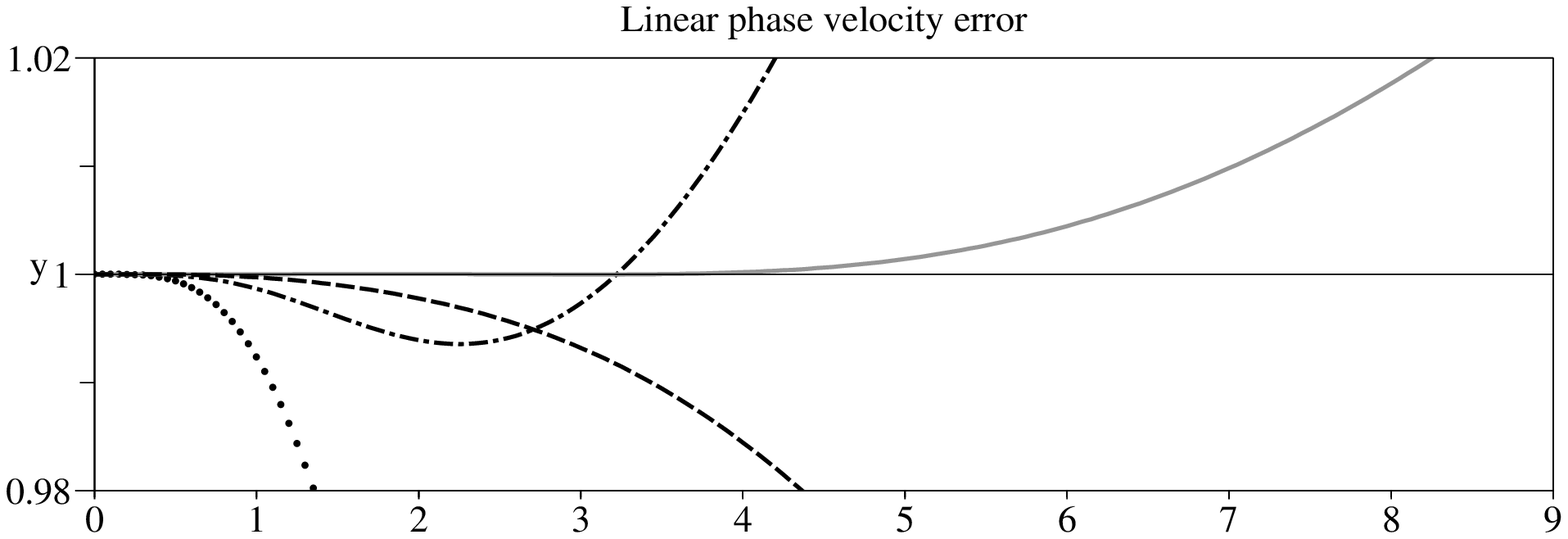}

\psfrag{y}[cr][cl][0.8]{$\dsp\frac{C_{GN}^{g}}{C_{S}^{g}}$}
\psfrag{Linear group velocity error}[0.7]{\hspace{7em}{\small Linear group velocity error}}
\hspace{0.5em}\includegraphics[width=0.985\textwidth]{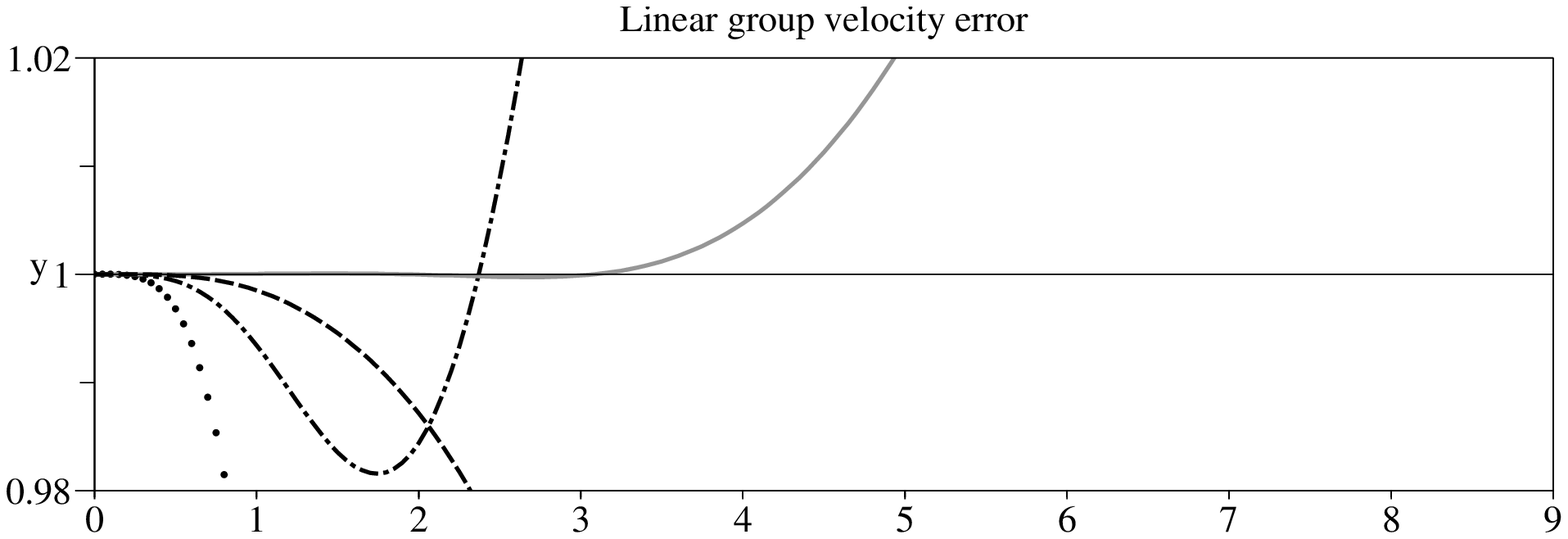}

\psfrag{x}{$kh_0$}
\psfrag{y}[cr][cl][1]{$\dsp\gamma_s$}
\psfrag{Linear shoaling coefficient}[0.7]{\hspace{7em}{\small Linear shoaling coefficient}}
\hspace{0.5em}\includegraphics[width=0.985\textwidth]{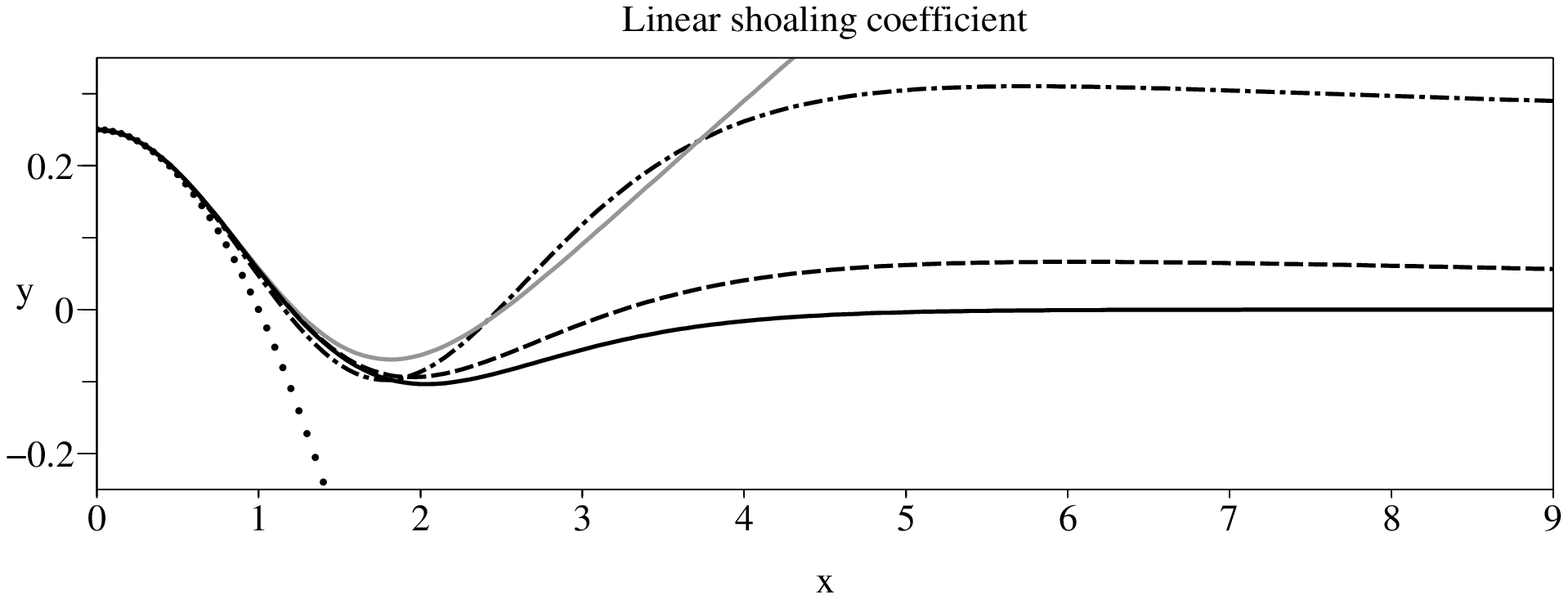}
\caption{Errors on linear phase (top) and group velocities (bottom), and shoaling coefficient. The reference from Stokes theory is in black plain line, the standard GN model ($\alpha = 1$) is in dots,  the optimized one-parameter model ($\alpha=1.159$) in dash-dots, the three-parameter model optimized for flat bottom (\ref{opt}) in grey plain line, and the three-parameter model optimized for uneven bottom (\ref{opt2}) in dashes.}
\end{figure}

As shown in Figure \ref{disp}, the dispersive range of validity of the model corresponding to the first set is considerably  extended (up to $kh_0=8$ for the phase velocity and up to $kh_0=5$ for the group velocity) but with deteriorated shoaling properties (satisfactory up to $kh_0 = 1.5$ only). On the other hand, the range of validity of $(\alpha_{opt}^{\,b},\theta_{opt}^{\,b},\gamma_{opt}^{\,b})$ model is similar to the one of the one-parameter model if we consider the phase and group velocities, but exhibits a very good shoaling up to $kh_0 =4$.

\section{Numerical scheme}\label{sectNS}

We present here the main lines of our numerical scheme. The splitting scheme is explained in \S \ref{sectsplit}; we then
show in \S \ref{secthyp} and \S \ref{sectdisp} how we treat respectively the hyperbolic and dispersive parts
of the equations. 

\subsection{The splitting scheme}
\label{sectsplit}

We decompose the solution operator $S(\cdot)$ associated to (\ref{GNgamma}) at each time step
by the second order splitting scheme
\begin{equation}\label{S0}
	S(\delta_t)=S_1(\delta_t/2)S_2(\delta_t)S_1(\delta_t/2),
\end{equation}
where $S_1$ and $S_2$ are respectively associated to the hyperbolic and dispersive parts of
the Green-Naghdi equations (\ref{GNdim}). More precisely:\\
$\bullet$  $S_1(t)$ is the solution operator
associated to the nonlinear shallow water equations
\be
	\label{S1_1D}
	\left\lbrace
	\begin{array}{l}
	 \dt h + \dx(hu)=0,\\
	 \dt(hu) + \dx\big( hu^2 + gh^2/2 \big) =-gh\partial_x b.
	\end{array}\right.
\ee
$\bullet$ $S_2(t)$ is the solution operator associated to the remaining (dispersive) part of the equations,
\be
	\label{S2_1D}
	\left\lbrace
	\begin{array}{l}
	 \!\dt h + \theta (1+\gamma h\cT\frac{1}{h})^{-1} \dx (h\cT u_\theta) = 0,\\
	\!\dt (h u)  - \frac{\theta+1}{\alpha+\theta} gh\dx\zeta
        +(1\!+\!(\alpha\!+\!\theta)h\cT\frac{1}{h})^{-1}\big[\frac{1+\theta}{\alpha\!+\!\theta} gh
        \dx\zeta+h\widetilde{\cQ}(u_\theta)\big]=0.
	\end{array}\right.
\ee
where the operators $\cT$ and $\widetilde{\cQ}$ are as in (\ref{defT}) and (\ref{Qbis}).

\medbreak

As far as the time discretization is concerned, we use
 a fourth-order explicit Runge-Kutta time scheme, both for the hyperbolic and the dispersive part.

\subsection{Spatial discretization of the hyperbolic part $S_1(\cdot)$}
\label{secthyp}

We use a high order finite volume approach in conservative variables $\vw = {}^t\left(h, hu\right)$, relying on Riemann problems for hyperbolic conservative laws \cite{godlewski_raviart}. This allows accurate computation of propagating bores, with reduced spurious effects of numerical dissipation and dispersion (this property is used in \cite{BCLMT} to handle wave breaking). Since we aim at computing the complex interactions between propagating waves and topography (including the preservation of motionless steady states), we embed this approach into a well-balanced scheme.

Based on discrete finite-volume cell averaging $\bar{\vw}_i^n$ at time $t^n=n\delta t$, we use $5^{th}$-order accuracy
WENO reconstructions, following \cite{jiang_shu}, together with the weight splitting method \cite{shi_shu}. Considering a cell $C_i$, this approach provides, for all $t^n$, interpolated quantities $\bar{\vw}_{i,l}$, $\bar{\vw}_{i,c}$ and $\bar{\vw}_{i,r}$, respectively at the left
boundary, center and right boundary of the cell.
To get a stable and well-balanced scheme, the following reconstructions are introduced, following \cite{audusse}:
$$
\left. \begin{array}{lll}
 b_i^*  & =  &\max(b_{i,r}, b_{i+1,l}),\\
h_{i,r}^* & =  &\max(0, h_{i,r} + b_{i,r} - b_i^*),\\
h_{i+1,l}^* &= &\max(0, h_{i+1,l} + b_{i+1,l} - b_i^*).
\end{array}\right.
$$
These left and right values for $h^*$ are used to provide auxiliary values $\vw_{i,r}^*$ and $\vw_{i+1,l}^*$:
\be
\vw_{i,r}^* = \left(\begin{array}{c}
h_{i,r}^*\\
h_{i,r}^* u_{i,r}
\end{array}\right),
\quad
\vw_{i+1,l}^* =\left( \begin{array}{c}
h_{i+1,l}^*\\
h_{i+1,l}^* u_{i+1,l}
\end{array}\right)
\ee
which are  injected into a Riemann solver. The corresponding semi-discrete finite-volume scheme for (\ref{S1_1D}) reads:
$$
\frac{d}{dt}\bar{\vw}_i (t) +   \frac{1}{\delta_x}   \Bigl( \Psi^r  \bigl(\bar{\vw}_{i,r},  \bar{\vw}_{i+1,l}, b_{i,r}, b_{i+1,l}
                    \bigr)   -  \Psi^l \left(\bar{\vw}_{i-1,r},\bar{\vw}_{i,l}, b_{i-1,r}, b_{i,l}  \right) \Bigr)=  S_{c,i}
$$
where $\Psi^r$ and $\Psi^l$ are numerical flux functions based both on a conservative flux consistent with the homogeneous SWE and the correction to the interface fluxes due to the hydrostatic reconstructions. $S_{c,i}$ is a centered discretization of the source term needed to achieve consistency and well-balancing properties. The reader is referred to \cite{audusse} for any details concerning the {\it hydrostatic reconstruction} and to \cite{berthon_marche} for details concerning the conservative flux, issued from a relaxation approach, used in our splitting method. To achieve an overall $4^{th}$ order accuracy, following \cite{noelle}, we define $S_{c,i}={}^t(0, S_{c,i}^{hu})$, with use of the following quadrature rule \cite{noelle}:
\begin{equation*}
S_{c,i}^{hu} =\frac{g}{6}\Bigl(4\bigl( (h_{i,l} + h_{i,c})( b_{i,l} - b_{i,c})  + (h_{i,c} + h_{i,l})( b_{i,c} - b_{i,l})  \bigr)   -   (h_{i,l} + h_{i,r})( b_{i,l} - b_{i,}) \Bigr).
\end{equation*}
This discretization allows both high-order accuracy and well-balancing properties. If the benefits are obvious concerning accuracy and convergence rates, the main drawback is the lack of robustness of this approach. In particular, we have to avoid any situation in which dry areas can occur, as in \cite{BCLMT}.

\subsection{Spatial discretization of the dispersive part  $S_2(\cdot)$}
\label{sectdisp}

As specified in \S \ref{sectsplit}, the system (\ref{S2_1D}) is solved at each time step using classical fourth-order accuracy finite-differences. For the sake of clarity, the reader is referred to \cite{BCLMT} for details concerning discretization of these dispersive terms, and also for the important issue of boundary conditions.

The choice of a finite difference method for solving $S_2$ also entails to switch between the cell-averaged values and the nodal values of each unknown, in a suitable way that preserves the global spatial order of the scheme.

\section{Numerical validations}\label{NV}

\subsection{Propagation of nonlinear cnoidal waves}
It has recently be shown in \cite{carter2010} that system (\ref{GNgamma}) with  $\alpha=1$, $\theta=\gamma=0$  admits the following family of periodic cnoidal  solutions:
\begin{subequations}\label{dnsoln}
\begin{equation}
h(x,t)=a_0+a_1\mbox{dn}^2\big{(}\kappa (x-ct),k\big{)},
\end{equation}
\begin{equation}
u(x,t)=c\Big{(}1-\frac{h_0}{h(x,t)}\Big{)},
\end{equation}
\begin{equation*}
\kappa=\frac{\sqrt{3a_1}}{2\sqrt{a_0(a_0+a_1)(a_0+(1-k^2)a_1)}},\;\;
c=\frac{\sqrt{ga_0(a_0+a_1)(a_0+(1-k^2)a_1)}}{h_0},
\end{equation*}
%\label{dnsoln}
\end{subequations}
where $k\in[0,1]$, $a_0>0$, $a_1>0$ are real parameters and $\mbox{dn}(\cdot,k)$ is a Jacobi elliptic function with elliptic
modulus $k$. The parameters of this solution can be related to physical variables in order to define (\ref{dnsoln}) in terms of wave
height $H$, wave period $T$, and mean water depth $h_0$. This can be achieved by solving the equations: 

\begin{subequations}
\begin{equation}
a_1=\frac{H}{k^2}, \;\;\;a_0=h_0-a_1 \frac{E(k)}{K(k)},\;\;\;
\hat{\omega}^2=\frac{3\pi^2 g a_1}{4\left[a_0 K(k) + a_1 E(k) \right]^2}
\end{equation}\end{subequations}
where $\hat{\omega}=2\pi/T$ is the angular frequency, 
$K(k)$ and $E(k)$ are the complete elliptic integrals of the first and second kinds respectively. \\
\begin{figure}[ht]
\begin{center}\includegraphics[width=0.8\textwidth]{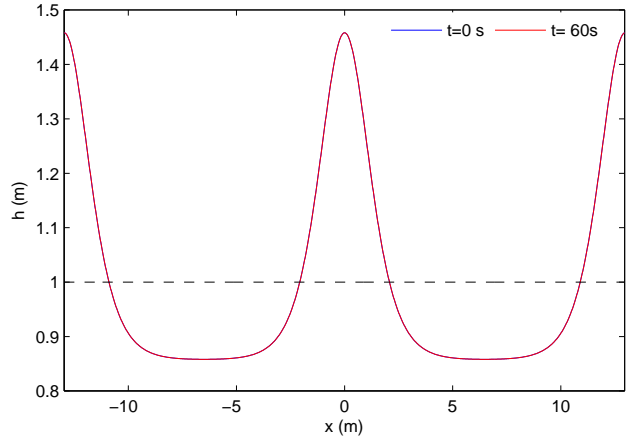}\end{center}
\caption{Propagation of a periodic cnoidal wave over a periodic domain ($H=0.6\,m$, $h_0=1\,m$, $T=4\,s$). Exact solution is ploted in blue line, numérical results at $t=60\,s$ in red line.} 
\label{fig:cnoidal}
\end{figure}
In this test case, we study the propagation of strongly non-linear cnoidal waves defined with $H = 0.6\,m$, $h_0 = 1\,m$ and $T = 4s$. The computational domain length is equal to $2$ wave-lengths and we use periodic boundary conditions. In Figure (\ref{fig:cnoidal}) numerical and theoretical solutions are compared after $15$ periods of propagation, at $t = 60s$. Computation is performed with $\delta x =$ and a Courant number equal to $1$. At $t = 60s$, we obtain a relative amplitude error of $1.3\, 10^{-3}\,\%$ and the relative celerity error is estimated to be less than $1.0\, 10^{-2}\,\%$.  Very accurate results are thus obtained, as an assessment of  the ability of our numerical method to compute in a stable way the propagation of non-linear cnoidal waves. 

\subsection{Propagation of highly dispersive waves}

\begin{figure}[ht]
\begin{center}\includegraphics[width=\textwidth]{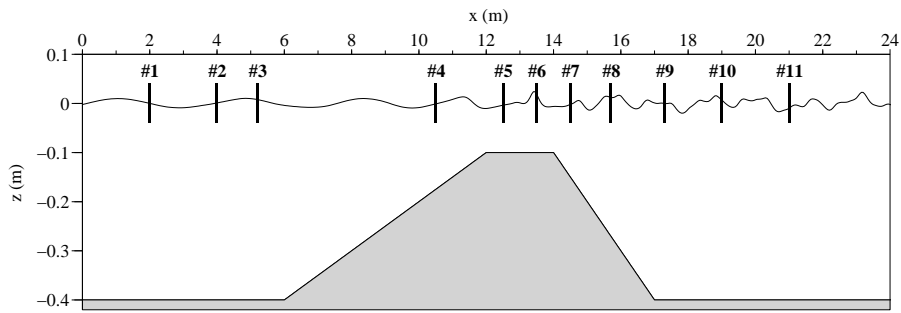}\end{center}
\caption{Experimental set-up and locations of the wave gauges} 
\label{fig:setup}
\end{figure}

In this section, we investigate the ability of the previous models to describe the propagation and the interaction of highly dispersive waves. To this end, we examine the propagation of regular waves over a submerged bar, using the set-up introduced by Beji \& Battjes (1993, \cite{Ding2}), and first used as a test by Dingemans (1994, \cite{Ding1}), in which the author evaluated the performance of various Boussinesq-type formulations by comparing computed free-surface time-series with experimental measurements at several gauges (see Figure \ref{fig:setup}). 

We consider here Case A of \cite{Ding1} where a small-amplitude long wave is generated at the left boundary : the free-surface elevation $a$ is $0.01m$, the time period $T$ is $2.02s$ and the initial depth $h_0$ is $0.4m$, which corresponds to a dispersion parameter $kh_0 \approx 0.7$. When the incident wave encounters the upward part of the bar, it shoals and steepens, which generates higher-harmonics as the nonlinearity increases. These higher-harmonics are then freely released on the downward slope, and become deep-water waves behind the bar. As discussed in Woo \& Liu (2001, \cite{Woo}), significant wave energy is present at $kh_0 \approx 4$ in the region behind the bar. For this reason, models based on a weakly dispersive assumption - such as Boussinesq-type ones or Green-Naghdi equations - are generally not able to correctly reproduce the measured profiles, since their linear dispersion properties usually become inaccurate beyond $kh_0 \approx 3$. As it involves highly nonlinear interactions and requires extended dispersive properties, this discriminating test has become a widely used benchmark over the past decade.

\begin{figure}[ht!]
% \begin{center}\hspace*{-2.5em}\includegraphics[width=1.21\textwidth]{test.eps}\end{center}
\begin{minipage}[t]{0.49\textwidth}
\begin{center}\includegraphics[width=\textwidth]{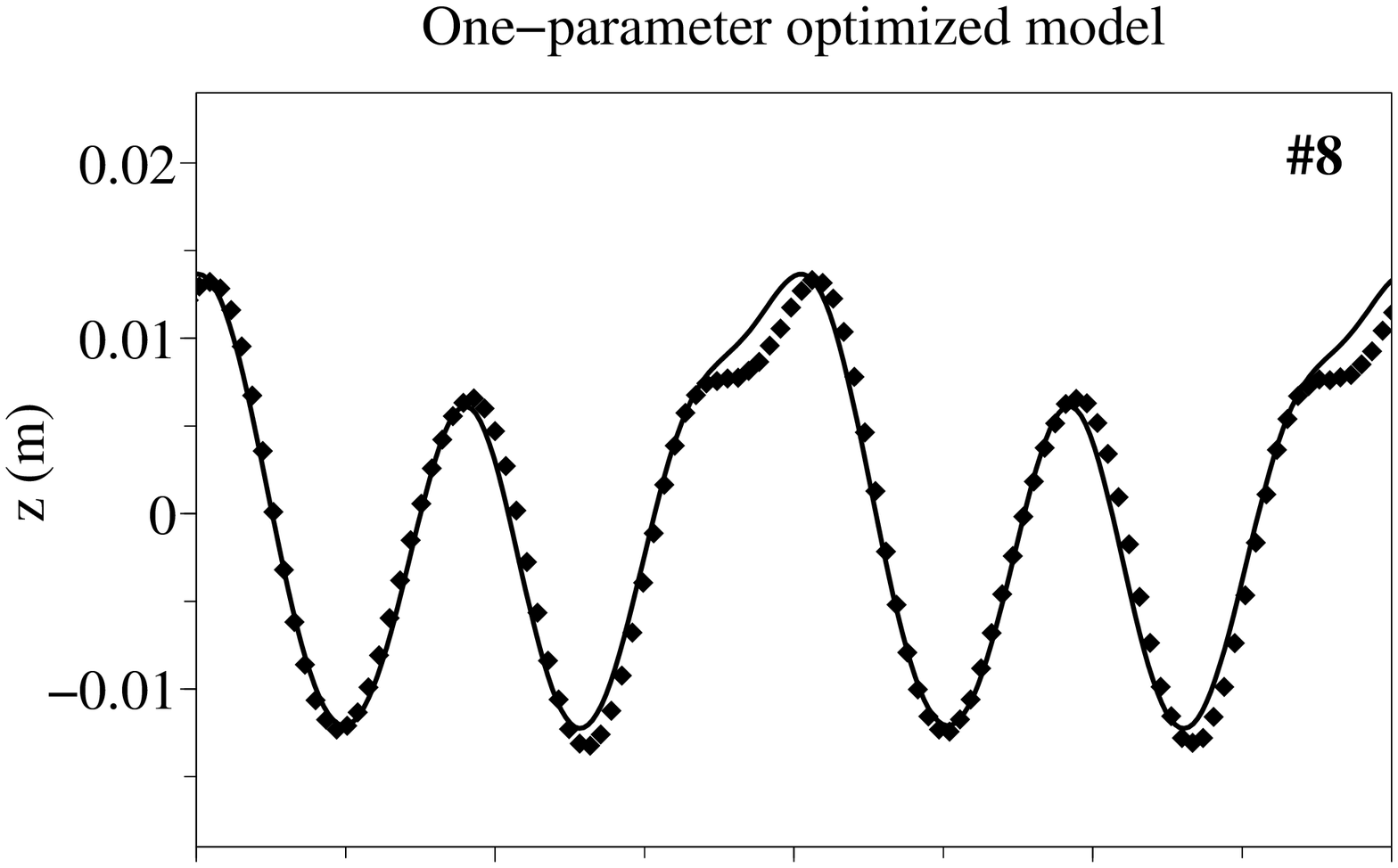}\end{center}
\vspace{-3em}
\begin{center}\includegraphics[width=\textwidth]{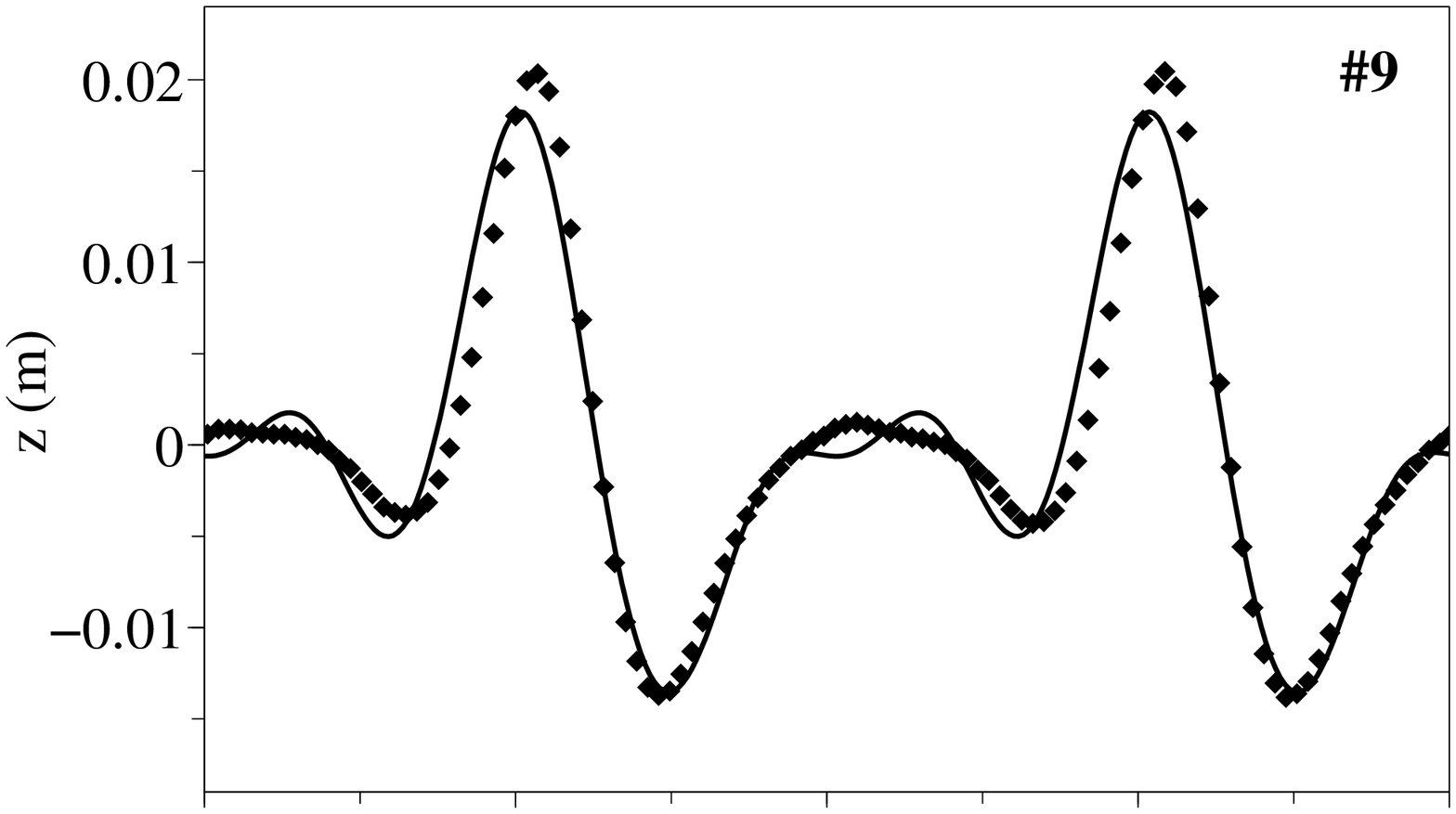}\end{center}
\vspace{-3em}
\begin{center}\includegraphics[width=\textwidth]{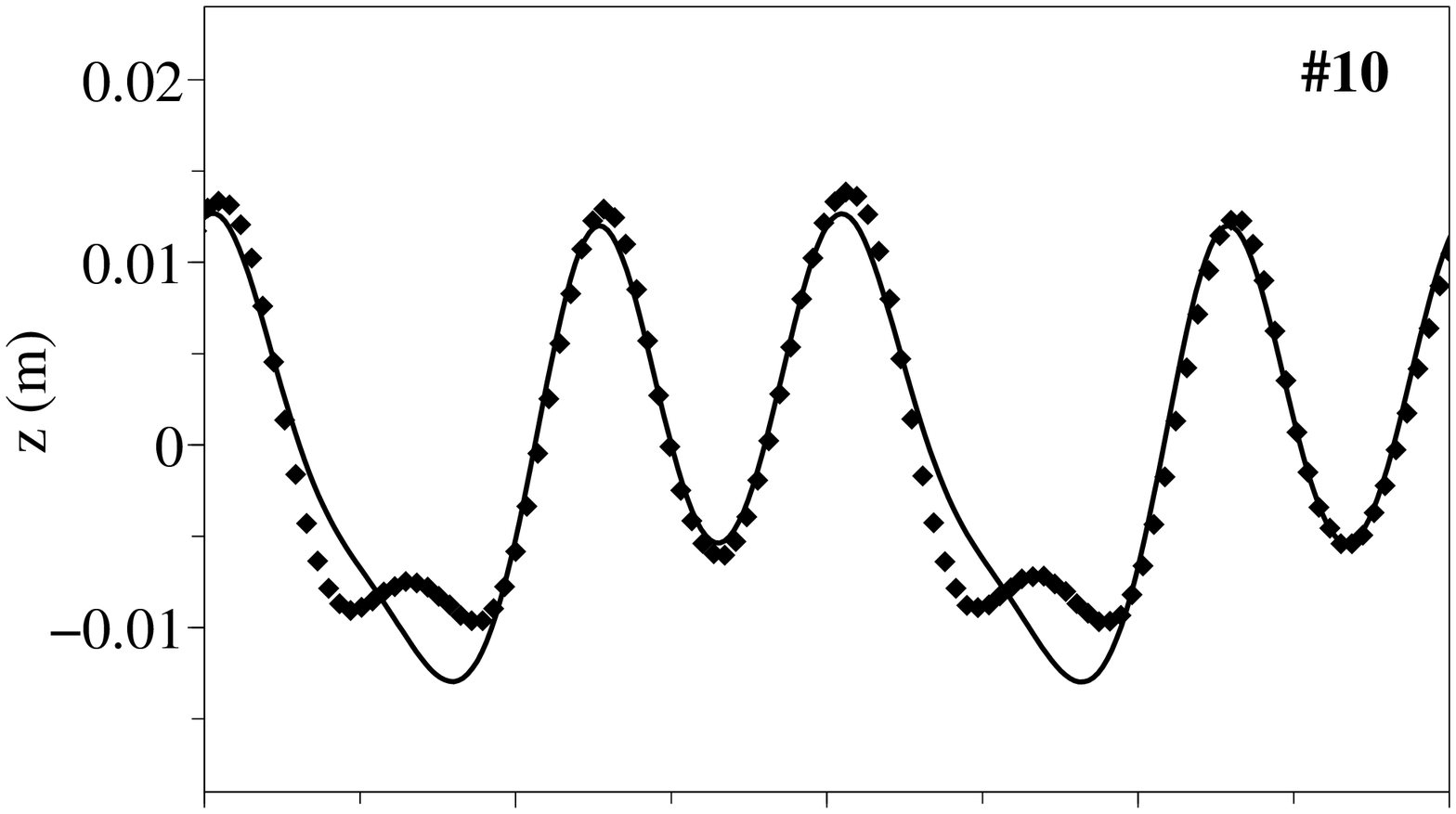}\end{center}
\vspace{-3em}
\begin{center}\includegraphics[width=\textwidth]{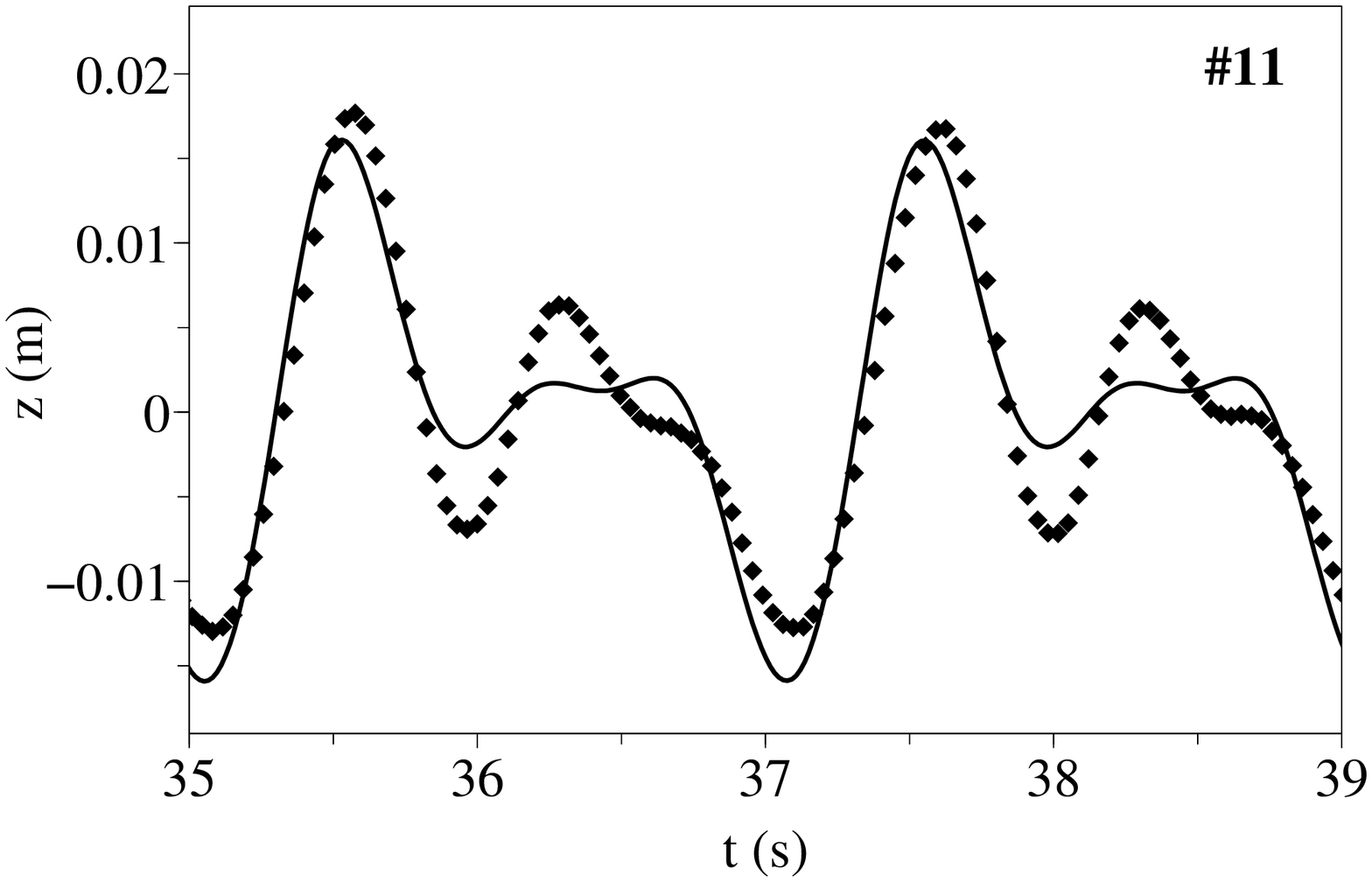}\end{center}
\end{minipage}
\begin{minipage}[t]{0.49\textwidth}
\begin{center}\includegraphics[width=\textwidth]{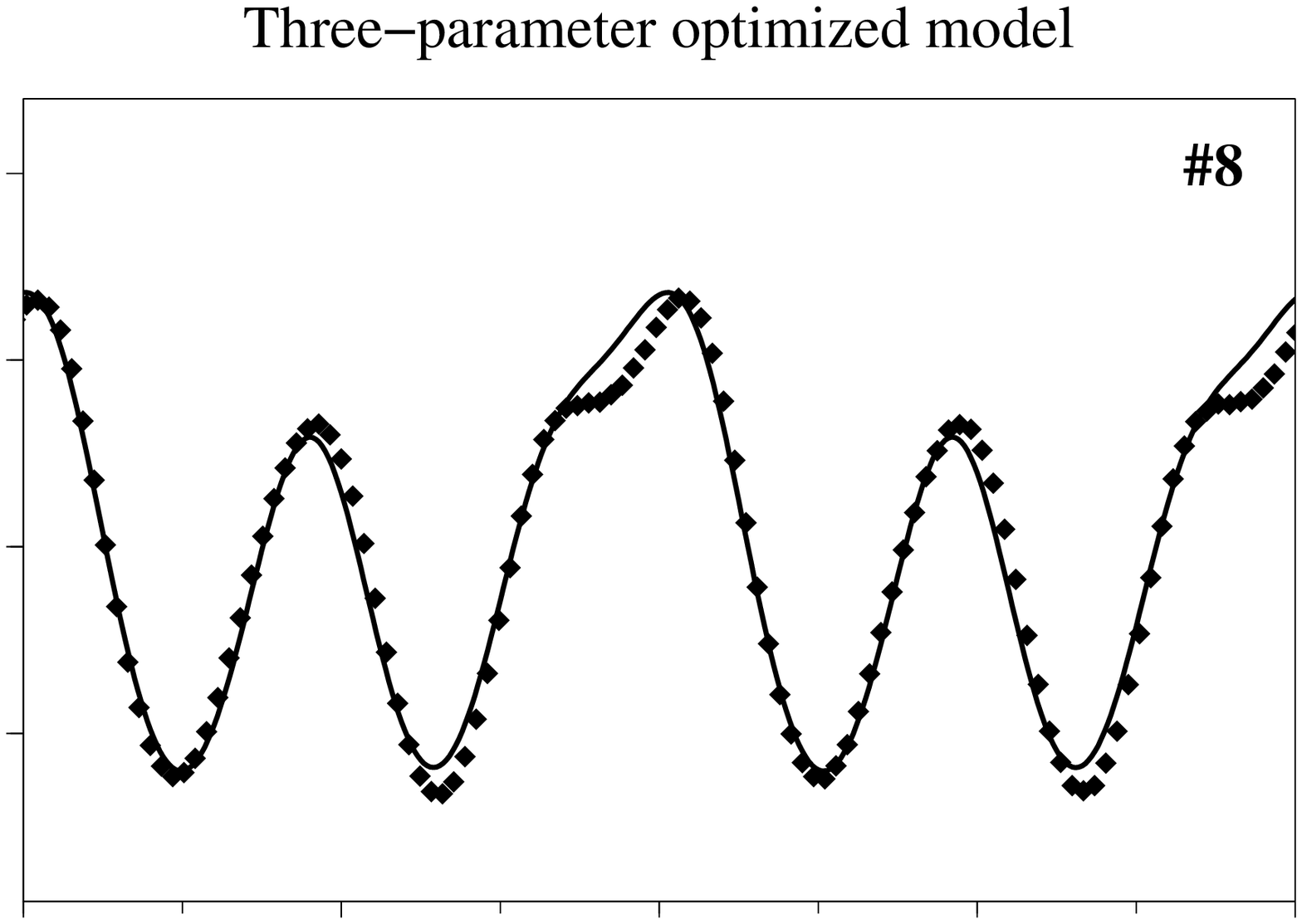}\end{center}
\vspace{-3em}
\begin{center}\includegraphics[width=\textwidth]{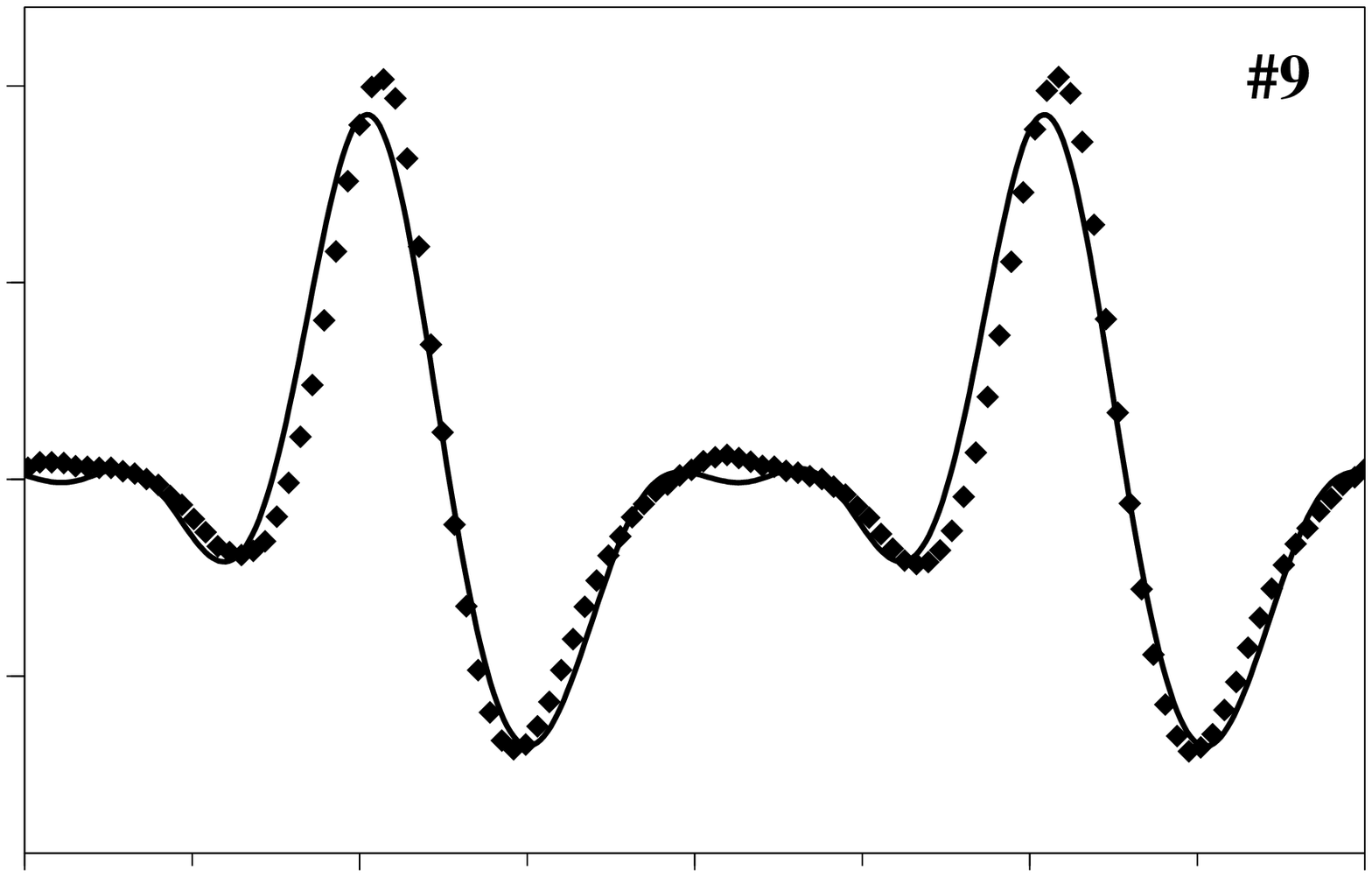}\end{center}
\vspace{-3em}
\begin{center}\includegraphics[width=\textwidth]{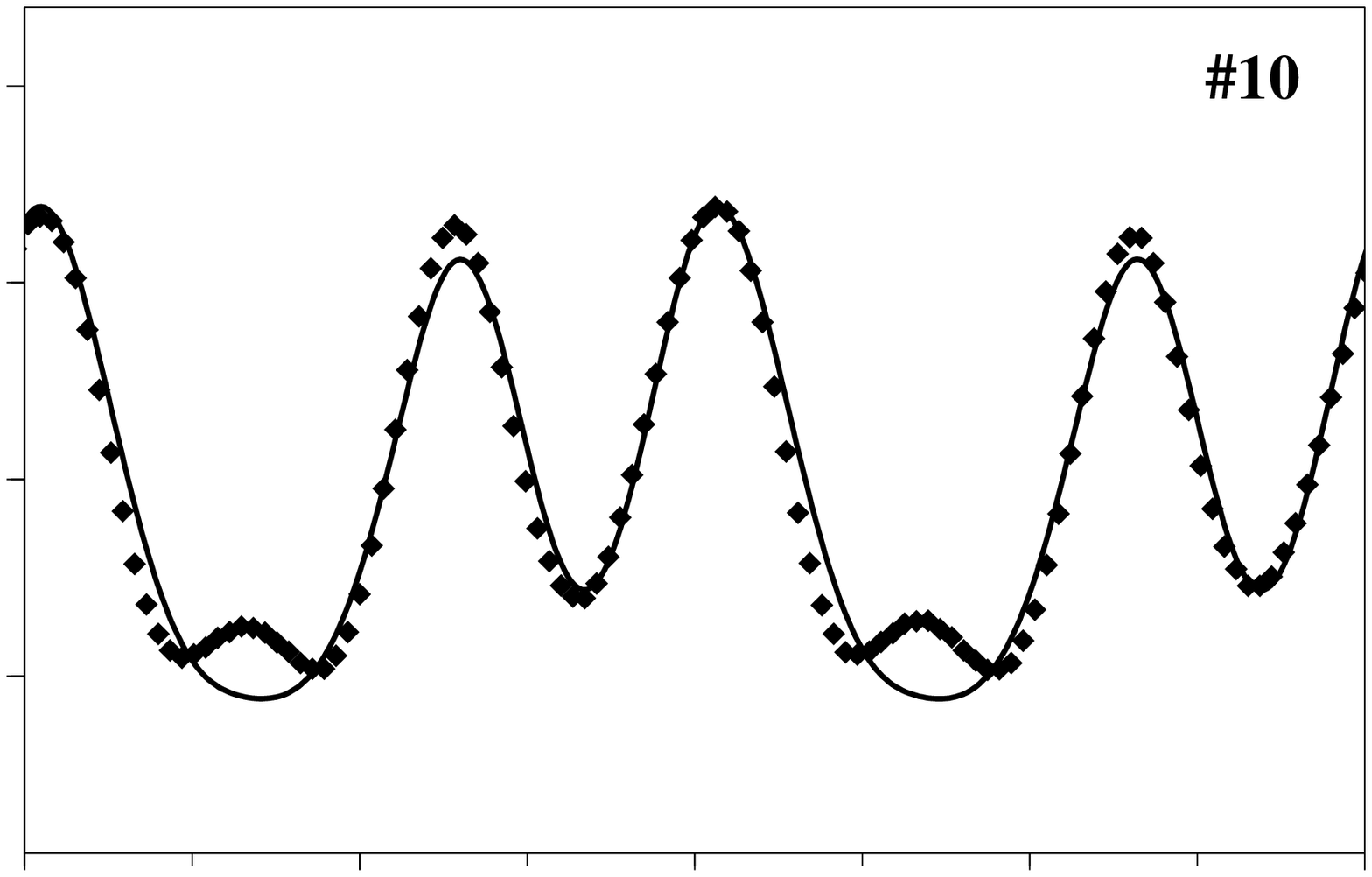}\end{center}
\vspace{-3em}
\begin{center}\includegraphics[width=\textwidth]{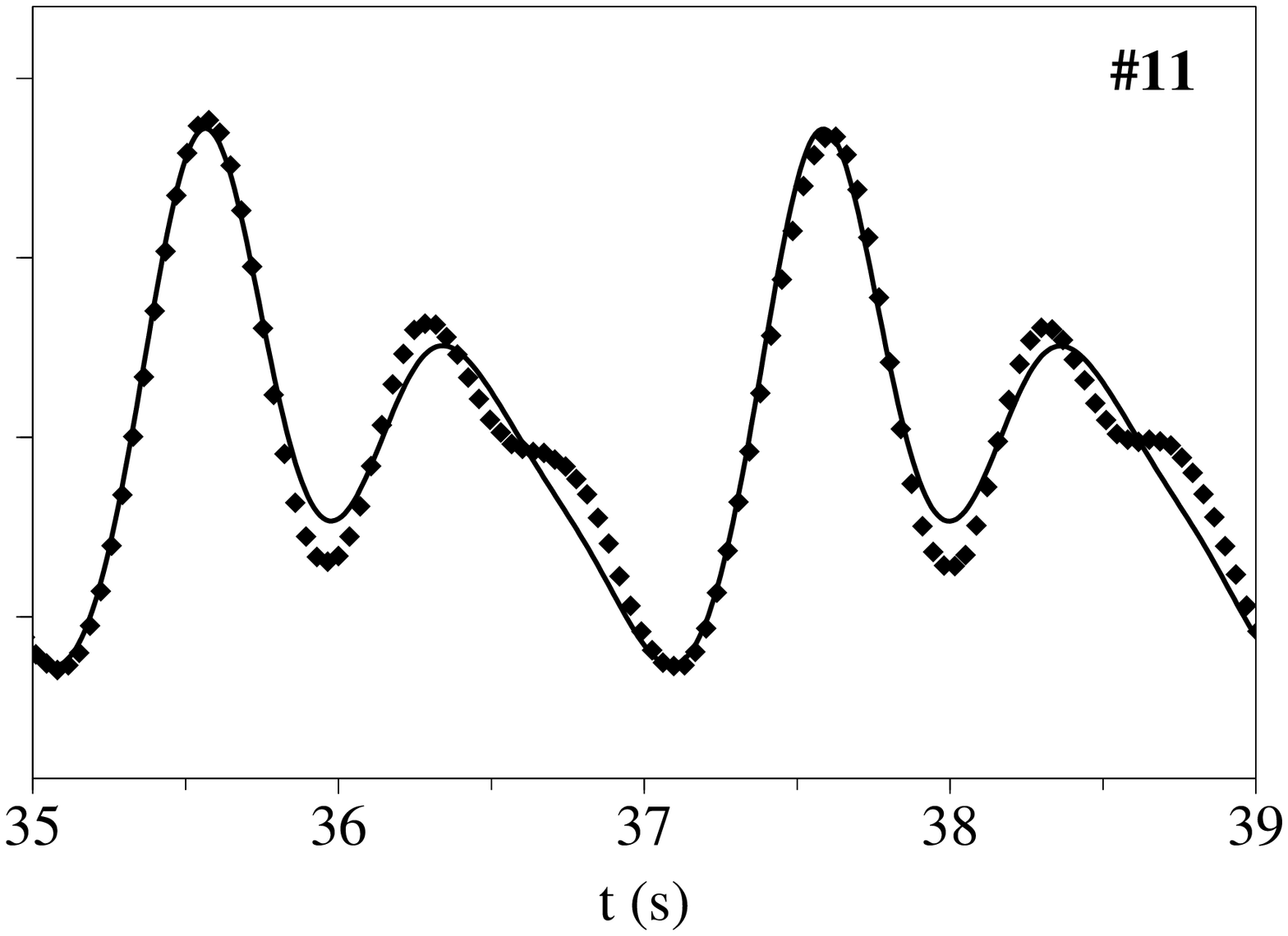}\end{center}
\end{minipage}
\caption{Comparison of computed free-surface time-series with experimental measurements for the optimized one-parameter Green-Naghdi model (left) and the three-parameter Green-Naghdi model optimized for uneven bottoms (right).}
\label{results}
\end{figure}

Numerical simulations for this case are shown in Figure \ref{results}, where computed free-surface time-series are compared to experimental measurements at the last four gauges, located on the downward part ($\#8$) of the bar and behind it ($\#9$, $\#10$ and $\#11$). Results at gauges $\#1$ to $\#7$ are not reported here as they do not exhibit any significant difference between computations and experimental data. We choose here to compare the numerical results of two Green-Naghdi models: the optimized one-parameter model (\ref{GNdim}) with $\alpha = 1.159$ and the optimized three-parameter model (\ref{GNgamma}) with $\alpha = 1$, $\theta = 0.207$ and $\gamma = 0.071$. We point out that the optimized three-parameter model for flat bottoms (\ref{opt}) yields results that are very similar to the optimized one-parameter model, and thus not reported here. For these simulations, we used the numerical parameters $\delta x = 0.03m$ and $\delta t = 0.01s$, which corresponds to a Courant number equal to $0.6$. The incident wave is generated at the left boundary using an absorbing-generating method presented in \cite{BCLMT}, and a sponge-layer is used at the right boundary to prevent perturbations by reflected waves. Computed time-series are shown over the time window $[35s,39s]$ where the waves kinematics are well established.

Both models show a similar and very satisfactory agreement with experimental measurements on the gauge $\#8$ located on the downward part of the bar, which is an expected result as the higher harmonics are not released yet. The first discrepancies between the two models appear on gauge $\#9$ located after the bar, where the three-parameter model provides a slightly better agreement with the measured profile than the one-parameter model. This remark persists on gauge $\#9$, even if both results still perform well. However, a very significant difference appear on the computed profiles at the very last gauge : the one-parameter model does not reproduce the correct profile (the amplitudes are either under-estimated or over-estimated) while the three-parameter profile shows an overall very good agreement with the results. This last gauge is by far the most discriminating one as the higher-harmonics are completely released at this location and can be seen as deep-water (and thus highly dispersive) waves. Keeping this in mind, the results obtained with our three-parameter model appears to be an excellent performance and clearly shows that the optimization presented in (\ref{opt2}) was very helpful in reproducing the correct profile. Moreover, seeing that the three-parameter model optimized for flat bottoms (\ref{opt}) yields less accurate results, even if this model exhibits significantly better dispersive properties on the linear and phase velocities, we can reasonably assume that the linear shoaling optimization is an unavoidable procedure for this very challenging test case. \\

\noindent
{\bf Acknowledgements} 

This work has been supported by the MathOcean project from the french Agence Nationale de la Recherche ANR-08-BLAN-0301-01.


\begin{thebibliography}{99}

\bibitem{AL} Alvarez-Samaniego, B. \& Lannes, D. 2007 Large time existence for 3D water-waves and asymptotics.
Invent. Math., {\bf 171}(3), 485--541.

\bibitem{audusse}  Audusse, E., Bouchut, F., Bristeau, M.-O., Klein, R. \& Perthame, B. 2004 A fast and stable well-balanced scheme with hydrostatic reconstruction for shallow water flows. {\it J. Comp. Phys.}, {\bf 25}(6), 2050--2065.

\bibitem{Ding2} Beji, S. \& Battjes, J. A. 1993 Experimental investigation of wave propagation over a bar. {\it Coastal Engineering}, {\bf 19}, 151--162.

\bibitem{berthon_marche} Berthon, C. \& Marche, F. 2008 A Positive Preserving High Order VFRoe Scheme for Shallow Water Equations: A Class of Relaxation Schemes. {\it SIAM J. Sci. Comput.}, {\bf 30}(5), 2587--2612.

\bibitem{BCLMT} P. Bonneton, F. Chazel, D. Lannes, F. Marche \& M. Tissier, {\it A splitting approach for the fully nonlinear and weakly dispersive Green-Naghdi model}, submitted.

\bibitem{carter2010} Carter, J.D. and Cienfuegos, R. 2010. Solitary and cnoidal wave solutions of the {S}erre equations and their stability. {\it Submitted to Physics of Fluids}. 

\bibitem{FC_DL} Chazel, F., Benoit, M., Ern, A. \& Piperno, S. 2009 A double-layer Boussinesq-type model for highly nonlinear and dispersive waves. \textit{Proc. R. Soc. Lond. A} \textbf{465}, 2319--2346.

\bibitem{cie2006} Cienfuegos, R., Barthelemy, E. \& Bonneton, P. 2006 A fourth-order compact finite volume scheme for fully nonlinear and weakly dispersive Boussinesq-type equations. Part I: Model development and analysis. {\it Int. J. Numer. Meth. Fluids}, {\bf 56}, 1217--1253.

\bibitem{cie2007} Cienfuegos, R., Barthelemy, E. \& Bonneton, P. 2007 A fourth-order compact finite volume scheme for fully nonlinear and weakly dispersive Boussinesq-type equations. Part II: Boundary conditions and validations. {\it Int. J. Numer. Meth. Fluids}, {\bf 53}, 1423--1455.

\bibitem{Ding1} Dingemans, M. W. 1994 Comparison of computations with Boussinesq-like models and laboratory measurements. {\it Report H-1684.12}, {\bf 32}, Delft Hydraulics.

%\bibitem{Ding3} Luth, H. R., Klopman, G., Kitou, N. 1994 Kinematics of waves breaking partially on an offshore bar; LDV measurements of waves with and without a net onshore current. {\it Report H-1573}, {\bf 40}, Delft Hydraulics.

\bibitem{godlewski_raviart} Godlewski, E. \& Raviart, P.-A. 1996 Numerical approximation of hyperbolic systems of conservation laws. {\it Applied Mathematical Sciences}, Vol. 118, Springer.

\bibitem{gre1976} Green, A. E., Naghdi, \& P. M. 1976 A derivation of equations for wave propagation in water of variable depth.  {\it J. Fluid Mech.}, {\bf 78}(2), 237--246.

\bibitem{jiang_shu} Jiang, G. \& Shu, C.-W. 1996 Efficient implementation of weighted ENO schemes, {\it J. Comp. Phys.} {\bf 126}, 202--228. 

\bibitem{shi_shu} Shi, J., Hu \& C., Shu, C.-W. 2002  A technique of treating negative weights in WENO schemes, {\it J. Comp. Phys. } {\bf 175}, 108--127. 

\bibitem{Lannes-Bonneton} Lannes, D. \& Bonneton, P. 2009 Derivation of asymptotic two-dimensional time-dependent equations for surface water wave propagation. {\it Physics of Fluids}, {\bf 21}, 016601.

\bibitem{GN_Grav} Le M\'etayer, O., Gavrilyuk, S., \& Hank, S. 2010 A numerical scheme for the Green-Naghdi model,  {\it J. Comp. Phys.}  {\bf 229}(6), 2034--2045. 

\bibitem{noelle} Noelle, S., Pankratz, N., Puppo, G. \& Natvig, J. 2006 Well-balanced ﬁnite volume schemes of arbitrary order 
of accuracy for shallow water ﬂows. {\it J. Comput. Phys.}, {\bf 213}, 47--499.

\bibitem{Nwogu} Nwogu, O. G. (1993) An alternative form of the Boussinesq equations for nearshore wave propagation. \textit{J. Waterway., Port, Coastal and Ocean Eng.} \textbf{119(6)}, 618--638.


\bibitem{sea1987} Seabra-Santos, F. J., Renouard, D. P. \& Temperville, A. M. 1987 Numerical and experimental study of the transformation of a solitary wave over a shelf or isolated obstacle. {\it J. Fluid Mech.}, {\bf 176}, 117--134.

\bibitem{sug1969} Su, C. H. \& Gardner, C. S. 1969 Korteweg-de Vries equation and generalizations. III. Derivation
of the Korteweg-de Vries equation and Burgers equation. {\it J. Math. Phys.} {\bf 10}(3), 536--539.

\bibitem{wei1995} Wei, G., Kirby, J.T., Grilli, S.T. \& Subramanya, R. 1995 A fully nonlinear Boussinesq model for surface waves. Part 1. Highly nonlinear unsteady waves. {\it J. Fluid Mech.}, {\bf 294}, 71--92.

\bibitem{Woo} Woo, S-B. \& Liu, P.L.-F. 2001 A Petrov-Galerkin finite element model for one-dimensional fully nonlinear and weakly dispersive wave propagation. \textit{Int. J. Num. Meth. Eng.} \textbf{37}, 541--575.

\end{thebibliography}
\end{document}